\documentstyle[preprint,prb,aps]{revtex}
 \title{Constrained equilibrium as a tool for characterization of 
 deformable porous media.
}

 \author{E.V. Vakarin$^a$, Yurko Duda$^b$, J.~P.~Badiali$^a$
}
 \address{
$^a$ UMR 7575 LECA ENSCP-UPMC, 11 rue P. et M. Curie, 75231 Cedex 05, Paris, 
France\\ 
$^b$ Programa de Ingenier\'ia Molecular, Instituto Mexicano del Petr\'oleo,
07730 D. F., M\'exico}

\begin{document}
\maketitle
\begin{abstract}
A new method for characterizing the deformable porous materials with
non-critical adsorption probes is proposed. The mechanism is based on a 
driving the adsorbate through a sequence of constrained equilibrium states
with the insertion isotherms forming a pseudo-critical point or a van der 
Waals-type loop. In the framework of a perturbation theory and Monte Carlo 
simulations we have found a link between the loop parameters and the host 
morphology. This allows one to characterize porous matrices through 
analyzing a shift of  the pseudo-critical point and a shape of 
the pseudo-spinodals.  
\end{abstract}


\section{Introduction}
One of the methods for characterization of porous materials 
\cite{review,Jaroniec} is a retrieving of pore size 
distributions through adsorption probes.
 Capillary condensation in macro- or meso-pores 
permits to detect them rather reliably because the condensation 
pressure and temperature are very sensitive to the pore size. Nevertheless, 
it becomes more and more questionable\cite{avalanches} that the true 
equilibrium behavior is observed in experiments. One of the major problems 
is a well developed metastability leading to very long equilibration times. 
Therefore, the condensation is usually detected as an onset of hysteresis. 
Nevertheless, it has recently been argued\cite{Kierlik2} that the hysteretic 
behavior is not necessarily associated with an underlying equilibrium phase 
transition. In the case of micro-pores the situation is even more difficult 
because such pores are too small to support any sharp (bulk-like) phase 
transition. This makes one to search for alternative 
characterization mechanisms which are not based on the critical behavior. 
In the case of deformable matrices\cite{PRB}, such as 
aerogels\cite{Pablo,Herman} or amorphous intercalation 
compounds\cite{BisqL}, the problem is additionally complicated by variations 
of the host morphology with the guest density or pressure. Thus the guest 
response (e.g., adsorption isotherms) is a superposition of at least two 
contributions - the adsorbate thermodynamics under the host confinement and 
the adsorbate-induced host deformation.

In this paper we propose a method for a  pore size 
determination through adsorption of non-critical fluids. The main idea
is to find an explicit relation between the density-dependent matrix 
morphology and some peculiarity in the fluid thermodynamic response.  
The mechanism is 
based on our finding\cite{JCPcomp} of negative compressibility 
states in confined dilatating systems. These states can appear as a 
consequence of a driving path - either a sequence of controlled lateral 
dilatations\cite{JCPcomp} or a sequential adsorption of prescribed 
portions\cite{Amarasekera}. This imposes an additional 
constraint\cite{Everett} on the fluid equilibrium. As a result, at certain 
conditions one can observe 
 a sequence of states which form a van der Waals-type loop in the isotherm. 
Then the loop parameters can be taken as the required specificities for a 
detection of the confining geometry. 

\section{Model}

The matrix is considered as a collection of slit-like pores with
different widths $h_i$ distributed according to some probability density.
For simplicity the pores are assumed to be uncorrelated. This allows us
to focus on a single pore of fluctuating width $h$ and surface area $S$. The 
width is known only statistically, that is, from a probability distribution
$f(h)$. The slit itself is a part of the hosting system, whose coupling
to the guest species changes the slit geometry (see below). 
The fluid is modeled as $N$ hard spheres of diameter $\sigma=1$,
injected into the pore. The hard core potential allows us to mimic
the excluded volume effects and ensures that the fluid is off-critical
(we do not consider freezing-type transitions).      

For any $h$ the total Hamiltonian is   
\begin{equation}
H=H_{ff}+H_{fw},
\end{equation}
where $H_{ff}$ is the hard sphere Hamiltonian, $H_{fw}$ is the slit
potential  
\begin{equation}
H_{fw}=A\sum_{i=1}^N 
\left[
\frac{1}{z_i^k}+\frac{1}{(h-z_i)^k}
\right];
\qquad
1/2 \le z_i \le h-1/2.
\end{equation}
We do not take into account a short-ranged fluid-wall attraction,
responsible for the surface adsorption or layering effects. The 
inverse-power shape for $H_{fw}$ is chosen as a generic form of a soft 
repulsion, with $k$ controlling the softness. In some sense this mimics
the hydrophobic or non-wetting effects that, as we will see later, are
essential. On the other hand, this
simple model takes into account that the walls are impenetrable and
ignores possible localized adsorption effects, which could complicate
the overall picture.   
For technical purposes we are 
working with $k=3$. Moreover, our results are qualitatively insensitive to a 
particular choice of $k$. 

In principle, one may consider two adsorption mechanisms. One is 
an equilibrium between the pore and the bulk fluids. In 
that case $N$ would change according to the chemical potential $\mu=\mu_b$. 
In this paper we treat another mechanism. Namely, an adsorption in 
doses\cite{Amarasekera}, when $N$ is injected in prescribed portions.    
In that case we have a sequence of the fluid equilibrium
states corresponding to different $N$. The sequence is determined 
by an injection procedure, which imposes an additional constraint.   
 It should be emphasized that 
choosing the model we tried to retain only a 
minimal number of relevant ingredients: the absence of liquid-gas transition 
in the fluid bulk, excluded volume effects, and a 
density-dependent heterogeneous confinement. 
\section{Perturbation theory}
\subsection{Conditional  insertion isotherm}
At any pore width $h$ the free energy can be represented as
\begin{equation}
\beta F(h)=\beta F_0(h)-\ln \langle e^{-\beta H_{fw}}\rangle_0
\end{equation}
where $F_0(h)$ is the free energy of a reference system, and 
$\langle...\rangle_0$ is the average over the reference state.  
Taking a spatially confined hard sphere system (the one 
with $A=0$) as a reference, we consider a first order 
perturbation\cite{JCPcomp} for the conditional free energy
\begin{equation}
\beta F(h)=\beta F_0(h) +\beta S \rho (h) \int \prod_{i} d{z}_i H_{fw} 
\end{equation}
where $\beta=1/(kT)$ and $\rho(h)$ is the pore density in the "slab"
approximation 
\begin{equation}
\label{rh}
\rho(h)=\frac{N}{S(h-1)},
\end{equation}
ignoring a non-monotonic behavior of the  
density profile with increasing pore density. This is reasonable for wide 
pores and low fluid densities. 
The reference part is estimated in the excluded volume approximation, while
the perturbation contribution is $N\Psi(h)$, such that the total conditional
free energy is
\begin{equation}
\label{Fh}
\beta F (h)=-N\ln{\left[
\frac{1-b\rho(h)}{\Lambda_T^3\rho(h)}
\right] 
} -N +N\beta\Psi(h)
\end{equation}
where $b=2\pi\sigma^3/3$ is the excluded volume factor, $\Lambda_T$ is
the thermal de Broglie length and
\begin{equation}
\label{Psi}
\Psi(h)=16A\frac{h}{(1-2h)^2}
\end{equation}

The conditional insertion isotherm can be calculated in the standard way   
\begin{equation}
\label{muh}
\beta \mu(h)=\left(\frac{\partial \beta F(h)}{\partial N} 
\right)_{\beta,S,h}=
\ln{\left[
\frac{\Lambda_T^3\rho(h)}{1-b\rho(h)}
\right] 
}+\frac{b\rho(h)}{1-b\rho(h)} + \beta\Psi(h)
\end{equation}
which relates the fluid density $\rho(h)$ and the chemical potential 
$\mu(h)$. 
\subsection{Matrix dilatation}
As is discussed above, the pore dilatation can be taken into account,
assuming that the width $h$ is density-dependent. Real insertion materials
are usually rather complex (multicomponent and heterogeneous). For that 
reason one usually deals with a distribution of pore sizes or with an
average size. In this context we assume that $h$ is known only statistically
and the probability distribution $f(h|\rho)$ is 
conditional\cite{JCPcomp} to the guest density $\rho$, which should 
be selfconsistently found from
\begin{equation} 
\label{rp}
\rho=\int dh f(h|\rho) \rho(h). 
\end{equation}
This relation reflects an injection sequence $\rho(h)$ and the matrix 
dilatation. Note that the average density $\rho$ is not uniquely determined
by the particles insertion but is also conditional to the matrix response.
Namely this constraint makes the problem non-trivial, inducing a coupling
between the insertion and the dilatation modes.  
Taking our result for $\mu(h)$, we can focus on the 
insertion isotherm, averaged over the width fluctuations.  
\begin{equation}
\label{mu}
\mu(\rho)=\int dh f(h|\rho) \mu(h).
\end{equation}
This, however, requires a knowledge on the distribution $f(h|\rho)$.
Even without resorting to a concrete form for $f(h|\rho)$, it is
clear that the matrix reaction can be manifested as a change in
the distribution width or/and the mean value.  
One of the simplest forms
reflecting  at least one of these features is a $\delta$-like 
distribution, ignoring a non-zero width: 
\begin{equation}
f(h|\rho)=\delta[h-h(\rho)],
\end{equation}
where $\delta(x)$ is the Dirac $\delta$-function and the mean pore width
$h(\rho)$ is density-dependent. For concreteness we consider the swelling
behavior\cite{BisqL}. 
Therefore, the mean pore width $h(\rho)$ increases with $\rho$. 
For illustration purposes we take the $S$-shaped form, considered 
previously\cite{PRB,JCPcomp} 
\begin{equation}
\label{hr} 
h(\rho)=h_0(1+\tanh[\Delta (\rho-\rho_0)]+\tanh[\Delta \rho_0]).
\end{equation}
This form resembles a non-Vegard behavior,
typical for layered intercalation compounds\cite{PRB}. 
The dilatation is weak at low densities 
($\rho << \rho_0$), the most intensive 
response is at $\rho \approx \rho_0$, and then the pore reaches a 
saturation, corresponding to its mechanical stability limit. Here  $\Delta$ 
is the matrix response constant or dilatation rate, controlling the 
slope near $\rho \approx \rho_0$, and $h_0$ is the mean pore width in the 
absence of insertion. This should be considered as a generic form that 
samples a non-linear increase of the average pore width, resolved at a given 
density $\rho$. The model mimics certain heterogeneity 
aspects which, as we believe, are essential. On the other hand, this choice 
allows us to map the problem in question onto the negative compressibility 
problem\cite{JCPcomp} in systems with fluctuating geometry.

From eq.~(\ref{rp}) the average density is found 
to be 
\begin{equation}
\rho=\frac{N}{S}\frac{1}{h(\rho)-1}
\end{equation}
Changing the surface density $N/S$ (by increments in $N$) we vary the 
average pore density $\rho$. This allows us to eliminate $N/S$ in the favor 
of $\rho$ in all thermodynamic functions. Combining eqs~(\ref{muh}), 
(\ref{rp}) and (\ref{mu}) we obtain 
\begin{equation}
\label{mur}
\beta \mu (\rho)=
\ln{\left[
\frac{\Lambda_T^3\rho}{1-b\rho}
\right] 
}+\frac{b\rho}{1-b\rho} + 
16\beta A\frac{h(\rho)}{(1-2h(\rho))^2}
\end{equation}
In the case of wide pores and a weak dilatation we expand $\beta \mu$ in 
terms of $1/h_0$ and $\Delta$, obtaining a generic van der Waals form  
\begin{equation}
\beta \mu (\rho)=
\ln{\left[
\frac{\Lambda_T^3\rho}{1-b\rho}
\right] 
}+\frac{b\rho}{1-b\rho} +
\frac{4\beta A}{h_0}  
-\frac{12\beta A}{h_0} \rho \Delta
\end{equation}
Therefore, as in the previous study\cite{JCPcomp}, we have an interplay of 
several effects -- the packing (first two terms), the fluid-matrix 
interaction  (density-independent), and the dilatation 
(linear in density). 
 It is seen that a coupling of the fluid-slit repulsion ($A$), spatial 
 confinement ($h_0$) and the dilatation ($\Delta$) acts 
 as an effective infinite-range fluid-fluid attraction. The latter induces
 an inflection point (or even a loop) in the isotherm $\mu(\rho)$. 
Introducing a dimensionless temperature $T^*=1/(4A^*)$, and solving
\begin{equation}
\frac{\partial (\beta\mu)}{\partial \rho}=0; \qquad
\frac{\partial^2 (\beta\mu)}{\partial \rho^2}=0
\end{equation} 
we find the pseudo-critical parameters at which the inflection point appears.
\begin{equation}
\label{parameters}
\rho_c=\frac{1}{3b}; \qquad
T^*_c=\frac{4}{27}\frac{\Delta}{bh_0}
\end{equation}
Plugging these back into eq.~(\ref{mur}) we get the third parameter
$\mu_c=\mu_c(b,h_0,\Delta)$. 

As has already been mentioned, the loop is not a signature of
a phase transition. It is a sequence of equilibrium states,
constrained by a coupling between the insertion and the dilatation modes. 
The physical mechanism is quite similar to the previously 
reported\cite{JCPcomp}. We deal with a competition of two effects:
the interparticle and wall-particle repulsions tend to increase the
insertion energetic cost $\mu$ with increasing $\rho$, while the pore 
dilatation diminishes it. As a result, at certain parameters the isotherm
becomes nonmonotonic. For a purely attractive pore ($A<0$) the inflection
point does not appear. Therefore, the wall-particle repulsion is an 
essential ingredient. This suggests that a suitable combination of the 
repulsive and attractive parts is a criterium for choosing the adsorption 
probe. As it  should be, the inflection (loop) feature 
disappears in the bulk limit $h_0 \to \infty$ or in the case of rigid 
matrices $\Delta \to 0$.  

In order to improve the accuracy of the theoretical results, the 
reference part is replaced by the Carnahan-Starling form 
\begin{equation}
\label{CS}
\beta \mu=\ln\left( \rho \Lambda_T^3\right)+
\frac{8\eta-9\eta^2+3\eta^3}{(1-\eta)^3}+
16\beta A\frac{h(\rho)}{(1-2h(\rho))^2};
\qquad
\eta=\pi \rho/6.
\end{equation}
This allows us to avoid the unphysical behavior at high densities and 
escape from the flaws of the excluded volume approximation (\ref{mur}).
In particular, all the pseudo-critical parameters ($\rho_c, T_c^*, \mu_c$)
appear to be related to the parameters of the pore width distribution
($h_0, \rho_0, \Delta$). Therefore, driving the system close to the 
pseudo-critical point permits to estimate the density dependent pore
size distribution. 
\section{Simulation}
In order to verify the existence of the loop predicted by the perturbation 
theory as well as to test the accuracy of the theoretical estimation, the 
simulation of the model in question has been carried out appling the Widom 
test particle insertion method \cite{Widom}. As in our previous 
work\cite{JCPcomp} we applied the conventional canonical $NVT$ MC 
simulations of a confined hard sphere fluid. The simulation cell was 
parallelepiped in shape, with parallel walls at surface separation $h$, and 
constant surface area $S=L_x \times L_y = 12\times12$. The periodic boundary 
conditions were applied to the $X$ and $Y$ directions of the simulation box; 
the box length in the $Z$ direction is fixed by the pore width. For a given 
pore width the adsorbed fluid density is chosen according to the 
Eq.(\ref{rh}) varying the number of particles, $N$. At frequent intervals 
during this simulation, a coordinate ${\bf r}_{N+1}$ has been generated 
randomly and uniformly over the simulation cell. For this value of 
${\bf r}_{N+1}$, we have computed 
$\langle \exp (-\beta H({\bf r}_{N+1})) \rangle_N 
$ averaged over all generated trial positions. In such a manner the chemical 
potential of the system with density $\rho$ has been calculated, 
\begin{equation}
\beta \mu = -\ln \left[ 
\frac{\rho}{\langle \exp (-\beta H({\bf r}_{N+1})) \rangle_N}
\right],
\end{equation}
where $H$ is the system Hamiltonian.
  
Each simulation runs $4\times 10^5$ MC cycles, whith the first 
half for the system to reach equilibrium 
whereas the second half for evaluating the ensemble averages.

\section{Results}
Adsorption isotherms are plotted in Figure~1. It is seen that the
inflection point around $\rho \approx 0.3$ transforms into the loop
with increasing repulsion $A$ (or decreasing temperature $T^*$).
As expected\cite{JCPcomp}, the perturbation theory underestimates the
magnitude of $A$ at which these effects appear. The loop gets sharper
with increasing dilatation rate $\Delta$ and becomes more pronounced
with decreasing pore width $h_0$. 
It is remarkable that the fitting on the pseudo-critical (inflection) point 
(part (a)) or on the loop (part (b)) allows us to find the set of the 
distribution parameters ($h_0$, $\Delta$, $\rho_0$) in a perfect agreement 
with the simulation. Attempts to fit under other criteria (e.g. 
the least-square deviations) give incorrect values of these parameters.

In agreement with our theoretical prediction (\ref{parameters}), the 
simulation results confirm that the pseudo-critical point (and the loop)
is reachable at any appropriate choice of the model parameters.  
This is in contrast to what was reported\cite{JCPcomp} for the negative 
compressibility states in the same model with lateral stretching 
(increasing $S$ while $N$ is fixed). 
There the loop has been found to appear 
only if the slit reaction $\Delta$ reaches some threshold value $\Delta^*$ 
which involves a combination of $h_0$, $\rho_0$ and $A$. In other words, 
the variation of the average density $\rho$ due to the transversal 
dilatation should dominate its variation, induced by the changes in the 
surface density $N/S$ due to the lateral stretch. This difference suggests 
that the driving path (namely, changing the fluid density by a lateral 
stretch or by a particle injection) is really important for this sort of 
systems. On the other hand, in the context of the present study, the fact 
that the pseudo-critical parameters do not involve any additional condition, 
is quite favorable for making a link between these parameters and the matrix 
morphology.

In Figure~2 the pseudo-critical parameters are analyzed. It is seen
that $\rho_c$ grows with the dilatation rate $\Delta$ almost linearly
exhibiting a characteristic kink. The latter signals a crossover from a weak 
to a strong dilatation regime. The kink shifts to lower $\Delta$ and becomes 
less pronounced with decreasing $h_0$ (that is why it is almost 
undetectable for $h_0=5$). At low $\Delta$ the pseudo-critical 
density is practically independent of $h_0$, while at high $\Delta$ it 
approaches $\rho_0$. In agreement with our simplified estimation 
(\ref{parameters}), $T_c^*(\Delta)$ can be reasonably approximated by a 
straight line (the lower inset). Nevertheless, it exhibits a weak inflection 
point at the same $\Delta$ (around $\Delta=4$) as $\rho_c$ does. The 
pseudo-critical temperature incresases with decreasing $h_0$. This is also 
coherent with the estimation (\ref{parameters}) predicting a decrease in the 
slope inversly proportional to $h_0$. In contrast, the insertion energy 
$\mu_c$ decreases and becomes insensitive to $h_0$ with increasing $\Delta$ 
(the upper inset). This agrees well with the physical intuition that the 
pore dilatation diminishes an overlap between the repulisive wall fields, 
making the insertion less energetically consuming.

In practice, an exact capture of the inflection point could be too 
demanding. Driving  at rather low temperatures ($T<T_c^*$), 
such that the system develops the loop, should be much easier. In that case 
a shape of the pseudo-spinodals could also be used as an indicator of the 
matrix morphology. These curves are plotted in Figure~3. It seen that
the pseudo-spinodals become broader with decreasing dilatation rate and   
shift to higher $\rho$ with increasing $\Delta$.
Comparing this with the previous figure, we can see again that $\rho_c$ 
converges to $\rho_0$ with increasing dilatation rate $\Delta$. 
If necessary, other quantitative characteristics, such as the curve 
asymmetry with respect to $\rho_c$ or $\mu$ values at the spinodal points, 
can also be easily extracted.

These results illustrate a correlation between the pseudo-critical point 
(or the loop) parameters and the density-dependent pore geometry.   
We did not try to obtain these parameters from the 
simulation because the procedure is time consuming, while new physical 
insights, compared to the anlytical estimations, are difficult to expect.  
The only promising point is to consider narrow ($1\le h_0 \le 2$) pores, 
for which the first-order perturbative scheme could fail even qualitatively. 
This point is left for a future analysis.

\section{Discussion and Conclusion}
A method for characterizing the deformable porous materials 
by adsorption of non-critical fluids is proposed. It is based
on a driving the host-guest system through a sequence of constrained
equilibrium states, forming a loop in the insertion isotherms. The 
constraint appears from a coupling of the insertion and the matrix
dilatation modes, inducing a competition between the repulsive
interactions and their relaxation by the host dilatation. As a result
the insertion isotherms develop pseudo-critical points or loops, whose
parameters are shown to be correlated with the parameters of the pore
size distribution. In particular, we have found that the magnitudes of the 
geometrical parameters ($h_0$, $\rho_0$, $\Delta$) which reproduce well the 
inflection point or the loop, perfectly agree with their values in the 
simulation. This suggests that the peculiar points (with $\partial 
\mu/\partial \rho=0$ or $\partial \rho$/$\partial \mu=0$) are extremely 
sensitive to the matrix morphology, while the overall isotherm shape is less 
important (unless on can achieve an exact isotherm coincidence). 
This is coherent with the recent finding\cite{Smit} that a matching
of the inflection points (with $\partial \rho$/$\partial \mu=0$) correctly 
reproduces the remaining part of the experimental isotherm. From the 
information-theoretical point of view\cite{power} such a criterium (the 
inflection matching) corresponds to a maximal amount of information one can 
get from a fitting of a model to a given experimental or simulation result.  
In our case we deal with the pseudo-critical point ($\partial \mu/\partial 
\rho=0$). Fitting on this peculiarity corresponds to a singular information 
rate\cite{power}, that allows us to update the geometrical parameters from 
their initial estimation in a limited range of $\rho$ (around the 
singularity). Namely this type of singularity is used to characterize
porous media\cite{review} through a shift of the fluid condensation
or freezing point.

Although we have taken a simplified $\delta$-like 
distribution, our conclusion should hold in general because the 
characteristic loop feature is directly related to the dilatation mode.
The latter translates into a shift of the mean pore size, that has been
taken into account. Nevertheless, the role of a non-zero distribution width 
remains to be investigated. The only restriction is that the number of the 
detected loop parameters should be (at least) equal to the number of 
parameters determining the pore size distribution. This restriction is not 
too strong because in practice the distribution is usually characterized by 
only few parameters (e.g. leading mean pore sizes and dipersions around 
them). On the other hand, the proposed method is mainly 
intended for a complementary detection of the porosity range (e.g. 
micropores) where other techniques fail. For illustration and technical 
purposes we were working with rather large pores $10 \le h(\rho) \le 30$. 
This, however, does not affect our main concept.   

It should be emphasized that, since 
we deal with a constrained equilibrium, the loop does not violate the 
thermodynamic stability\cite{Everett}. Therefore, this feature can be 
readily found under suitable experimental conditions, similar to those
realized in the low-pressure adsorption into microporous 
matrices\cite{Amarasekera}. 
In this respect it is worth noting that quite similar loop
features have been detected\cite{Amarasekera} for nitrogen and argon
adsorption in microporous carbons and synthetic zeolites.    
As in our case, the fluids have been inserted in sequential doses.
Nevertheless, in contrast to our assumption, there is no evidence
for an insertion-induced dilatation of these slit-shaped materials. 
The loops have been interpreted as a signature of a monolayer-induced
micropore filling\cite{Amarasekera}. In this context we can propose
an alternative interpretation of our results, making a link to the 
experimental observations. Namely, it is well known\cite{Amarasekera} that 
the micropore filling starts from the narrowest pores ($h_0$ in our laguage).
As the fluid pressure (or density) increases, the larger micropores 
progressively fill. Microscopically this effect appears from an interplay
of repulsive and attractive parts of the wall-particle interaction. The
latter depends on the chemical nature of the probe and the matrix geometry
and composition. In order to extract general features, one may accept this
effect as an experimental fact, trying to see its consequences.  

Therefore, in a given density domain the fluid
resolves a quite narrow range of pore sizes because the narrow pores are 
already blocked by a preceding dose, but the filling of larger pores is not 
thermodynamically favorable at this density. This fact is captured by
our distribution $f(h|\rho)$ that can be re-interpreted as an effective
density-dependend matrix heterogeneity, with $\Delta$ being the 
heterogeneity index. In other words, an increase of the mean pore size
$h(\rho)$ may result from a pore blocking effect.   
This interpretation is coherent with the one proposed
by the authors of \cite{Amarasekera}, who claimed that the S-shaped feature
(the loop) signals a switching from the ultramicropore to micropore 
filling. Nevertheless, this qualitative analogy between the matrix swelling 
and heterogeneity requires further analysis, including a competition of the 
repulsive and attractive interactions. This issue is left for a future study.


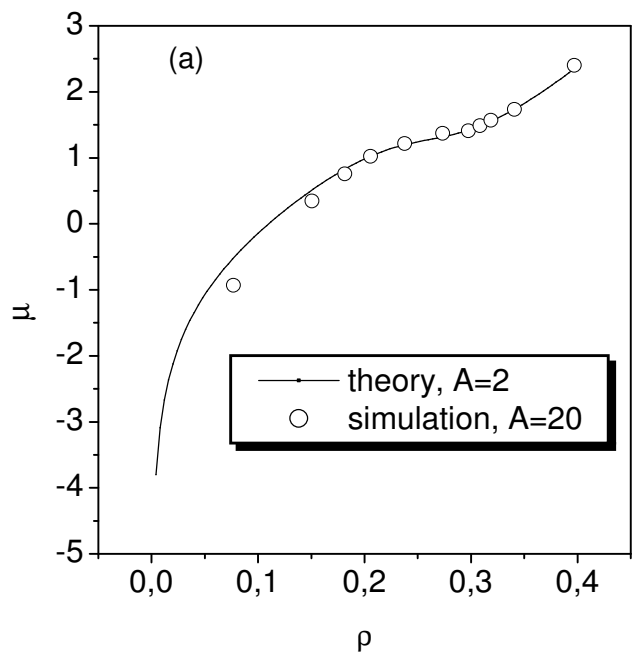
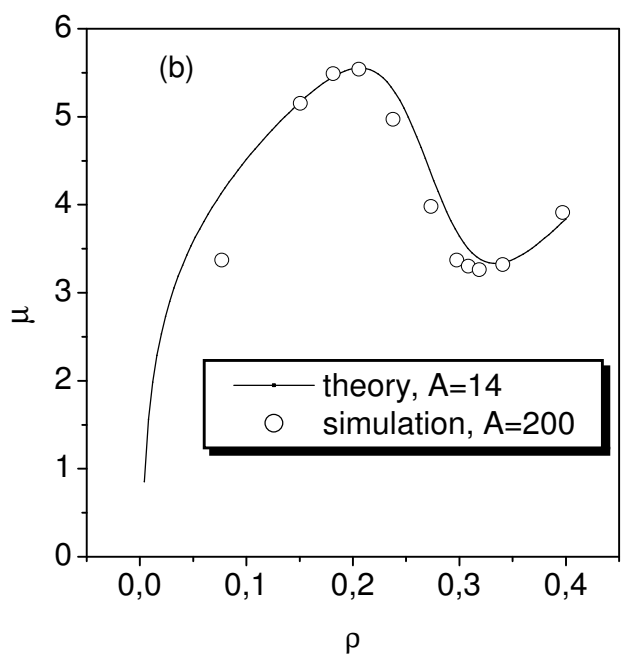
\begin{figure}
\caption{Insertion isotherms for
a weakly (a) and strongly (b) repulsive slit. The other parameters are
$h_0=10$, $\Delta=15$, $\rho_0=0.3$ (both theory and simulation)} 
\end{figure}
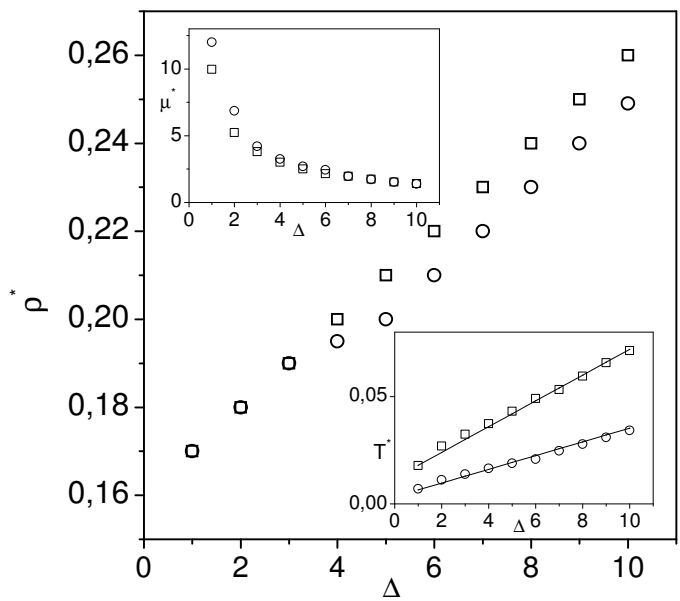
\begin{figure}
\caption{Variation of the pseudo-critical parameters with the dilatation 
rate $\Delta$. $\rho_0=0.3$, $h_0=5$-squares, $h_0=10$-circles 
 } 
\end{figure}
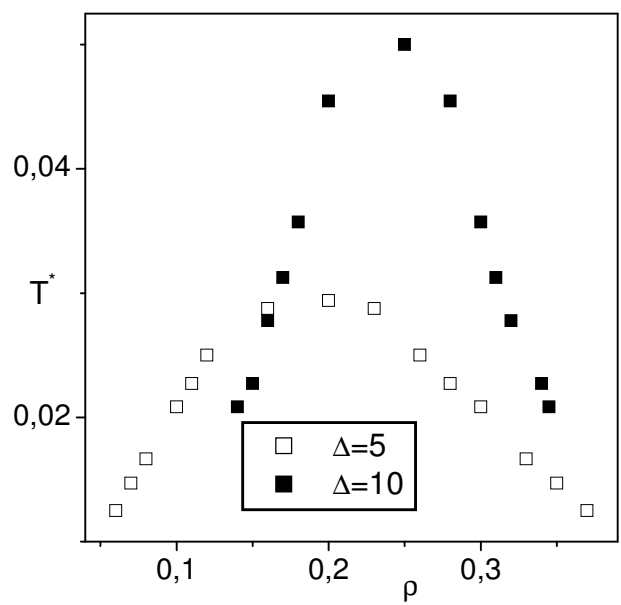
\begin{figure}
\caption{Pseudo-spinodal curves at different dilatation rates $\Delta$.
The other parameters are $h_0=7$, $\rho_0=0.3$} 
\end{figure}

 \end{document}